\documentclass[showpacs,prl,twocolumn,floatfix,
nofootinbib,superscriptaddress]{revtex4}

\usepackage{hyperref}
\hypersetup{
    bookmarks=true,         
    unicode=false,          
    pdftoolbar=true,        
    pdfmenubar=true,        
    pdffitwindow=false,     
    pdfstartview={FitH},    
    pdftitle={My title},    
    pdfauthor={Author},     
    pdfsubject={Subject},   
    pdfcreator={Creator},   
    pdfproducer={Producer}, 
    pdfkeywords={keyword1} {key2} {key3}, 
    pdfnewwindow=true,      
    colorlinks=true,       
    linkcolor=blue,          
    citecolor=blue,        
    filecolor=blue,      
    urlcolor=blue           
}

\usepackage{graphicx}
\usepackage{nicefrac}
\usepackage{lipsum}
\usepackage{amsmath,bm}
\usepackage{subfig}
\usepackage{multirow}

\newcommand\T{\rule{0pt}{2.6ex}}

\begin{document}


\title{Standard Model predictions for  $B \to K \ell^+ \ell^-$ with form factors from lattice QCD}

\author{Chris Bouchard}
\thanks{bouchard.18@osu.edu}
\affiliation{Department of Physics,
The Ohio State University, Columbus, OH 43210, USA}

\author{{G.\ Peter} Lepage}
\affiliation{Laboratory of Elementary Particle Physics,
Cornell University, Ithaca, NY 14853, USA}

\author{Christopher Monahan}
\affiliation{Physics Department,
College of William and Mary, Williamsburg, Virginia 23187, USA}

\author{Heechang Na} 
\affiliation{Argonne Leadership Computing Facility,
ANL, Argonne, IL 60439, USA}

\author{Junko Shigemitsu}
\affiliation{Department of Physics,
The Ohio State University, Columbus, OH 43210, USA}

\collaboration{HPQCD Collaboration}
\noaffiliation
\date{\today}


\begin{abstract}
We calculate, for the first time using unquenched lattice QCD form factors, the Standard Model differential branching fractions $d \mathcal{B}/dq^2(B\to K\ell^+\ell^-)$ for $\ell = e, \mu, \tau$ and compare with experimental measurements by Belle, BABAR, CDF, and LHCb.  We report on $\mathcal{B}(B \to K \ell^+ \ell^-)$ in $q^2$ bins used by experiment and predict $\mathcal{B}(B \to K \tau^+ \tau^-) = (1.44 \pm 0.15) \times10^{-7}$.  We also calculate the ratio of branching fractions $R^\mu_e = 1.00023(63)$ and predict $R^\tau_\ell = 1.159(40)$, for $\ell=e,\mu$.
Finally, we calculate the ``flat term" in the angular distribution of the differential decay rate $F_H^{e,\mu,\tau}$ in experimentally motivated $q^2$ bins.
\end{abstract}


\maketitle


\section{ Introduction }
The rare semileptonic decay $B \to K \ell^+ \ell^-$ is a $b \to s$ flavor-changing neutral current process that only occurs through loop diagrams in the Standard Model, making it a promising probe of new physics.  To make predictions for Standard Model observables, or extract information about potentially new short distance physics, knowledge of associated hadronic matrix elements is required.  Because hadronic matrix elements quantify nonperturbative physics, the only first-principles method for calculating them is lattice QCD.  
Hadronic matrix elements for semileptonic decays are parameterized by form factors.  
For processes that occur readily in the Standard Model only the vector and scalar form factors $f_{+,0}$ are phenomenologically relevant.  The study of rare decays requires knowledge of the tensor form factor $f_T$ as well.  All form factors are potentially important in the presence of new physics.

There is an active effort ~\cite{Becirevic:2012, Bobeth:2011, Beaujean:2012, Altmannsofer:2012, Bobeth:2013} to constrain new physics using experimental results for $B\to K\ell^+\ell^-$, often in combination with other rare $B$ decays.
In the past, the needed form factor information for these works has come from light cone sum rules ({\it cf.} Refs.~\cite{Ball:2005, Khodjamirian:2010, Khodjamirian:2013}), valid at low $q^2$.  In a more first principles approach, Ref.~\cite{Becirevic:2012} calculates the form factors in lattice QCD at high $q^2$ using the so-called quenched approximation~\cite{quench}, then extrapolates to low $q^2$ using the model-dependent BK parameterization~\cite{Becirevic:2000}.
However, given the number and precision of recent experimental measurements of this decay, and the importance of stringent tests of the Standard Model in such rare processes, Standard Model predictions free of uncontrolled approximations have become crucial.  In the lattice approach, for instance, it is imperative to go beyond the uncontrolled quenched approximation.  In this letter we present Standard Model results that are based for the first time on unquenched lattice calculations that take effects of up, down, and strange sea quarks into account.  Furthermore, our results are extrapolated over the full kinematic range of $q^2$ in a model independent way.  We then make detailed comparisons of these new Standard Model predictions with experimental measurements at the $B$-factories Belle~\cite{Wei:2009} and BABAR~\cite{Lees:2012}, by CDF~\cite{Aaltonen:2011}, and most recently by LHCb~\cite{Aaij:2012, Aaij:2012b}.
We note that there are preliminary unquenched lattice QCD results for the form factors by Liu {\it et al.}~\cite{Liu:2011} and the Fermilab Lattice and MILC collaboration~\cite{Zhou:2012}.

\section{ Lattice QCD calculation }
\label{sec-LQCD}
We begin with an overview of the lattice QCD calculation of the form factors $f_{0,+,T}$.  Ref.~\cite{Bouchard:PRD} contains details, provides the information required to reconstruct the form factors, and calculates useful ratios of form factors.  

Ensemble averages of two and three point correlation functions are performed using a subset of the MILC $2+1$ asqtad gauge configurations~\cite{Bazavov:2010}.  
We use two lattice spacings, $a\approx 0.12$ fm and $0.09$ fm, to allow extrapolation to the continuum and simulate at multiple light sea-quark masses to guide a chiral extrapolation to physical light-quark mass.
The valence quarks in our simulation are NRQCD $b$ quarks~\cite{Lepage:1992, Na:2012} and HISQ light and strange quarks~\cite{Follana:2007, Na:2010, Na:2011}.
Data were generated using local and smeared $b$ quarks, U(1) random wall sources for the light and strange valence quarks, and four values of momenta to guide the kinematic extrapolation.
We generate three point data for several temporal spacings between the $B$ meson and kaon to improve our ability to extract three point amplitudes.

We extract hadronic matrix elements from simultaneous fits to two and three point data using Bayesian fitting techniques~\cite{Lepage:2002} and incorporate correlations among data for different matrix elements and at different momenta.  
Effective vector and tensor lattice currents are matched to the continuum using one loop, massless-HISQ lattice perturbation theory~\cite{Monahan:2013}.  

We extrapolate to physical light quark mass and zero lattice spacing using fit ans\"atze based on partially quenched staggered chiral perturbation theory~\cite{Aubin:2007}.  
The extrapolations include NLO chiral logs, NLO and NNLO chiral analytic terms to accommodate effects of omitted higher order chiral logs, and finite volume effects.
We neglect the $\mathcal{O}(a^2)$ taste-breaking discretization effects in~\cite{Aubin:2007}, but accommodate generic discretization effects through $\mathcal{O}(a^4)$ in the extrapolation, including light- and heavy-quark mass-dependent discretization effects.

Using the physical extrapolated results we generate synthetic data for each form factor, restricted to the region of $q^2$ for which simulation data exist.  We extrapolate these data over the full kinematic range of $q^2$ using the model-independent $z$ expansion~\cite{Boyd:1996, Arnesen:2005} with the Bourrely, Caprini, and Lellouch (BCL) parameterization~\cite{Bourrely:2010}.  

\begin{figure}[t!]
{\scalebox{0.98}{\includegraphics[angle=0,width=0.5\textwidth]{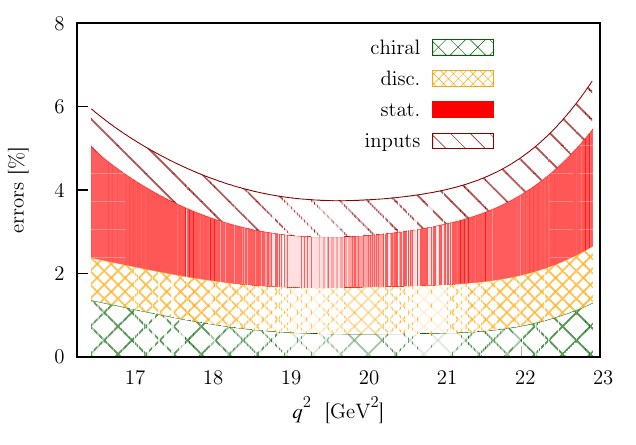}}}
\\
\vspace{-0.1in}
{\scalebox{0.98}{\includegraphics[angle=0,width=0.5\textwidth]{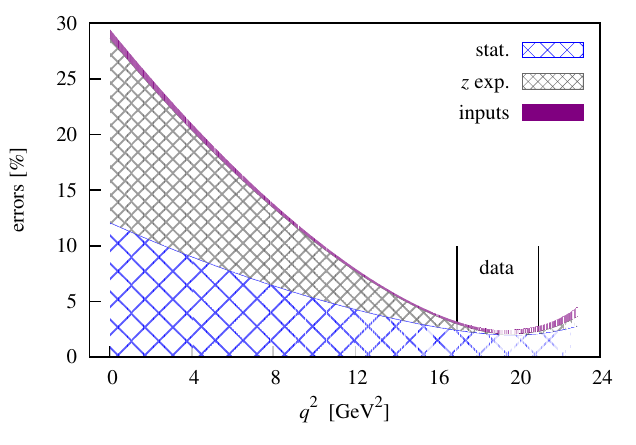}}}
\vspace{-0.25in}
\caption{Errors for $f_+$ from ({\it top}) chiral/continuum and ({\it bottom}) kinematic extrapolations.  
The total \% error is the sum in quadrature of kinematic extrapolation errors.}
\label{fig-FFerrors}
\end{figure}
The chiral/continuum extrapolation errors for $f_+$ are shown in the top plot of Fig.~\ref{fig-FFerrors} in the region of $q^2$ for which simulation data exist.  
Following the method outlined in~\cite{Davies:2008}, the error is separated into components by grouping related fit parameters.  
The chiral extrapolation error (``chiral'') contains errors in $f_+$ due to extrapolating to physical light quark mass, strange quark mass interpolations to correct slight mistunings, small contributions from mass differences due to the use of a mixed action (asqtad sea and HISQ valence quarks), and finite volume effects.  
Discretization errors (``disc.'') include light- and heavy-quark mass-dependent, and mass-independent discretization errors.  
Statistical errors (``stat.") represent the errors associated with the form factors obtained from the correlation function fits, {\it i.e.} the data for the chiral/continuum extrapolation fits.  
Errors due to input parameters are labeled ``inputs''.

Components of the kinematic extrapolation error are plotted in the bottom panel of Fig.~\ref{fig-FFerrors}, where the region of $q^2$ for which simulation data exist is indicated on the plot.
The ``stat." error is associated with the synthetic data generated by the chiral/continuum extrapolation, the ``$z$ exp." error is the sum in quadrature of errors from coefficients of the $z$ expansion, and the ``inputs'' error is from uncertainty in input parameters.

In the region of simulated $q^2$ the dominant source of error is from the synthetic data.  At low $q^2$ the error is roughly split between the synthetic data and the kinematic extrapolation.  Errors associated with input parameters are negligible.  A similar analysis of $f_{0,T}$ in Ref.~\cite{Bouchard:PRD} reveals similar behavior.

\begin{figure}[t!]
{\scalebox{0.98}{\includegraphics[angle=-90,width=0.5\textwidth]{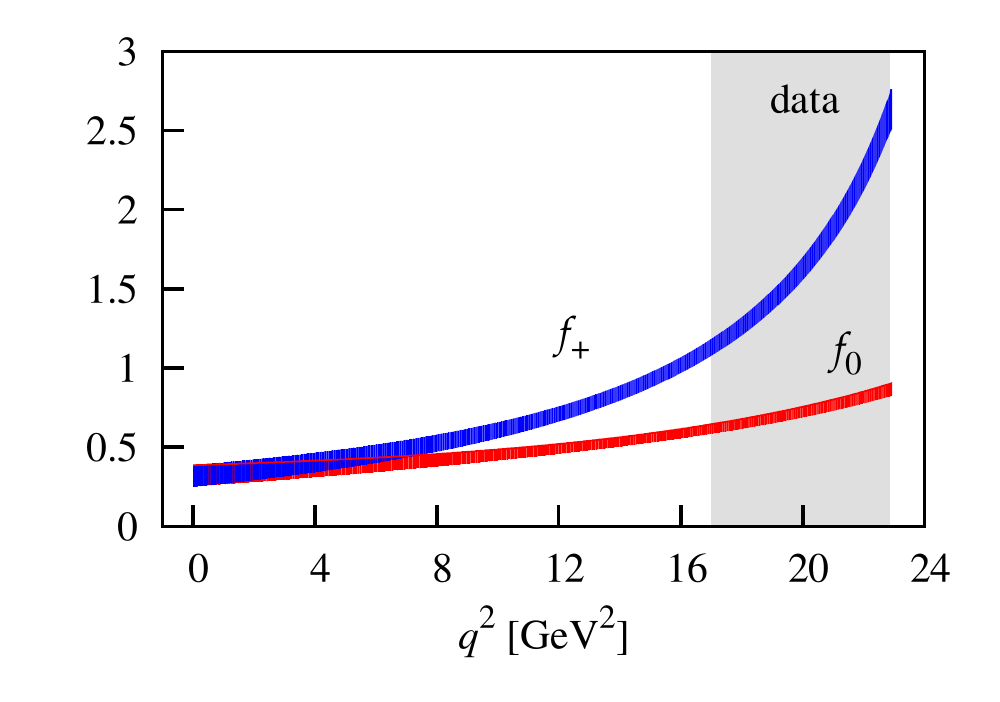}}}
\\
\vspace{-0.1in}
{\scalebox{0.98}{\includegraphics[angle=-90,width=0.5\textwidth]{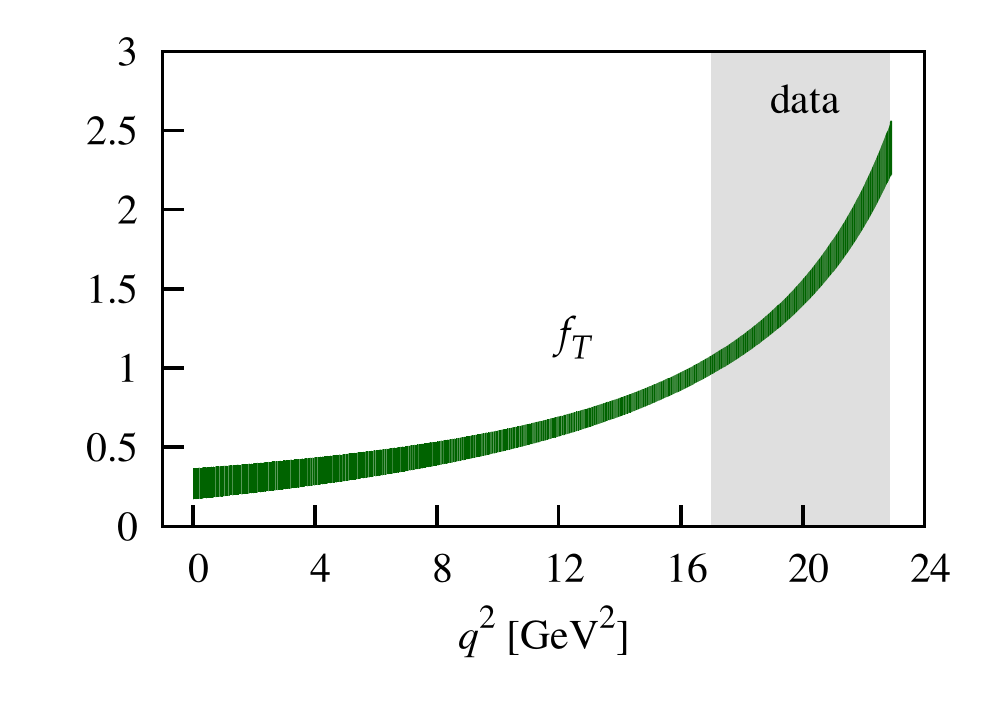}}}
\vspace{-0.15in}
\caption{Form factors for $B\to K\ell^+\ell^-$.}
\label{fig-finalFFs}
\end{figure}

In addition to fit errors, systematic errors from matching, electromagnetic and isospin breaking effects, and omission of charm sea-quarks contribute a combined 4\% error (dominated by matching).  The form factors, including all sources of error, are shown in Fig.~\ref{fig-finalFFs} with shaded bands indicating the region of simulation data.

\section{ Standard Model Observables }
%
\begin{figure*}[t!]
\vspace{-0.25in}
\hspace{-0.07in}  
\subfloat[][\label{fig-dBFdq2l} Belle~\cite{Wei:2009}, BABAR~\cite{Lees:2012}, CDF~\cite{Aaltonen:2011}, and LHCb~\cite{Aaij:2012, Aaij:2012b} data and the Standard Model contribution to $d\mathcal{B}_\ell / dq^2$, $\ell = e, \mu$.]
{\scalebox{0.98}{\includegraphics[angle=-90,width=0.5\textwidth]{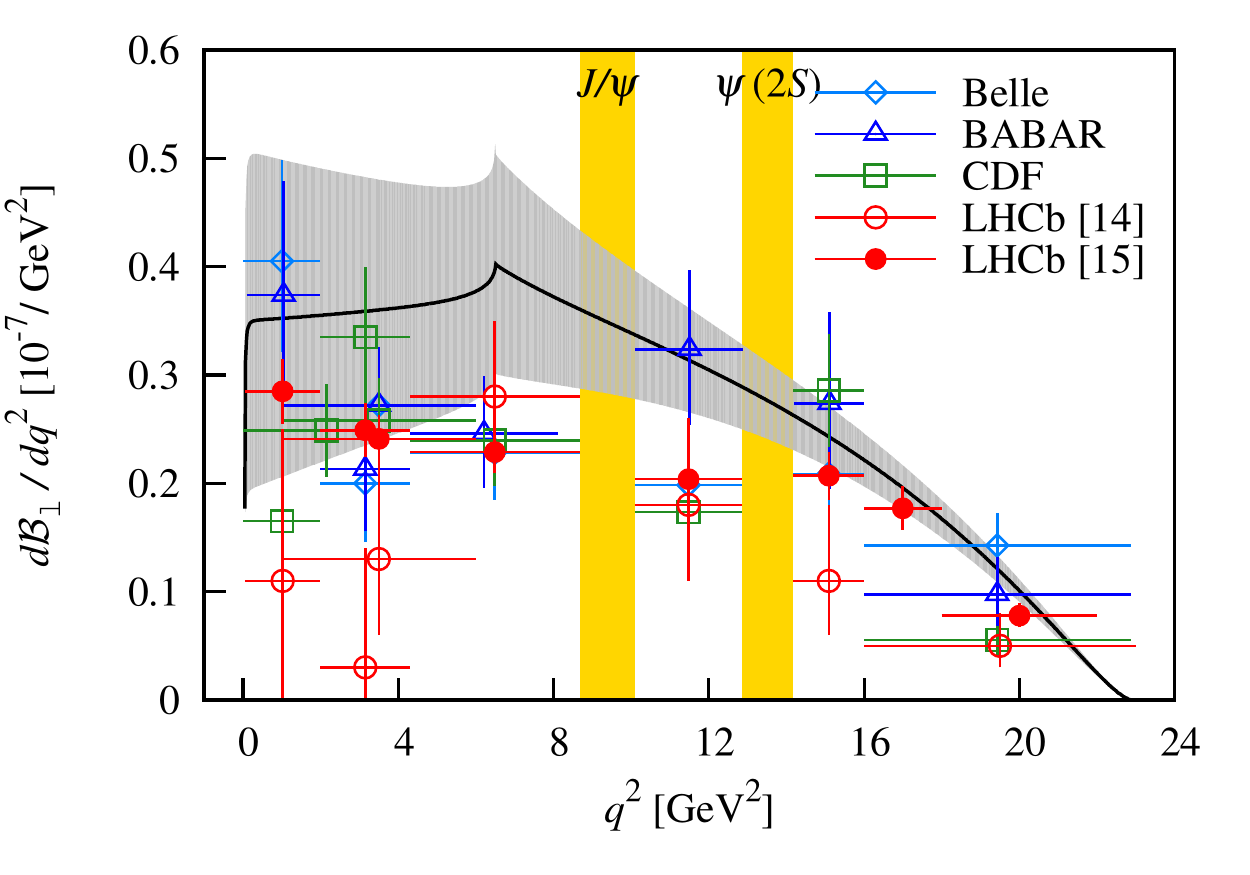}}}
\subfloat[][\label{fig-dBFdq2l_err}Error components for $d \mathcal{B}_\ell / dq^2 $.]
{\scalebox{0.98}{\includegraphics[angle=-90,width=0.5\textwidth]{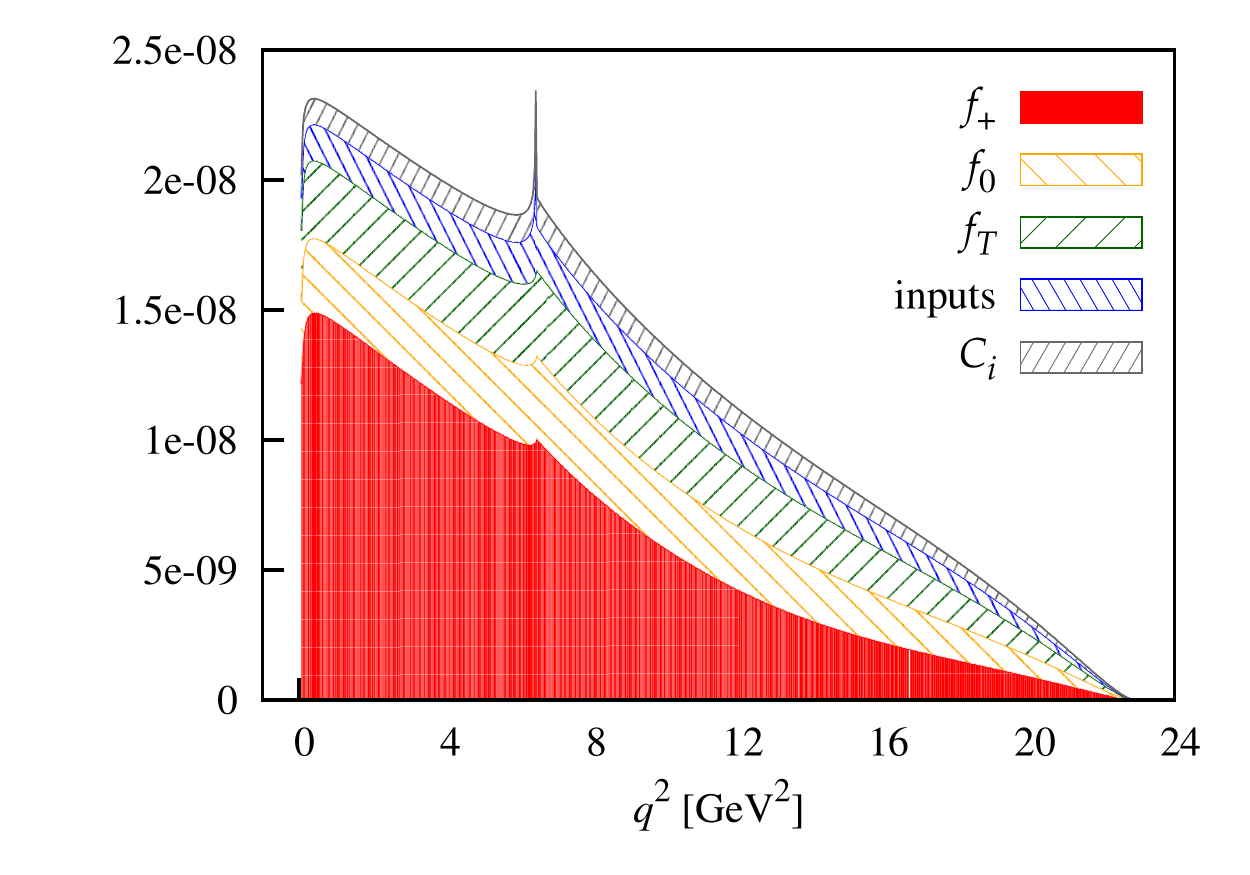}}}
\\
\subfloat[][\label{fig-dBFdq2tau}Predicted Standard Model contribution to $d \mathcal{B}_\tau / dq^2 $.]
{\scalebox{0.98}{\includegraphics[angle=-90,width=0.5\textwidth]{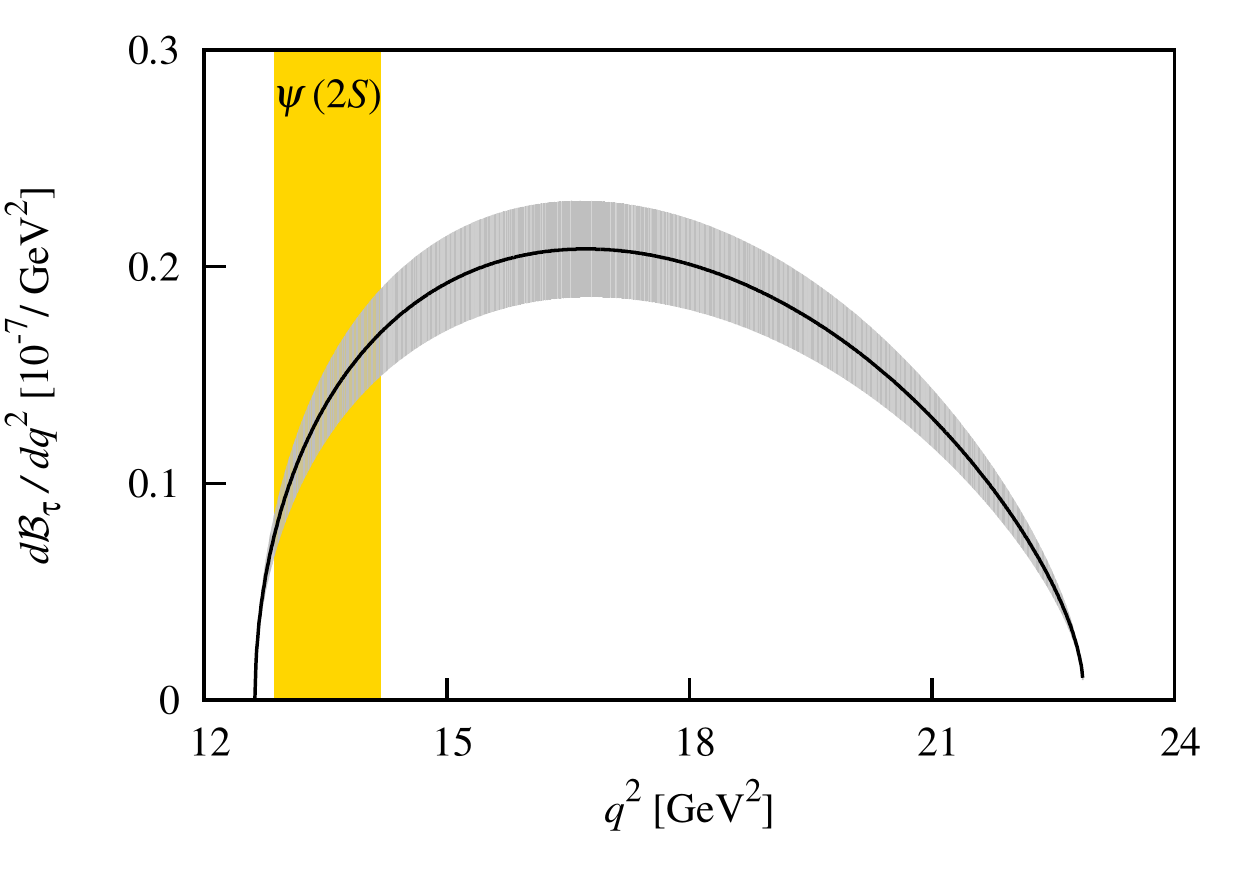}}}
\subfloat[][\label{fig-dBFdq2tau_err}Error components for $d \mathcal{B}_\tau / dq^2 $.]
{\scalebox{0.98}{\includegraphics[angle=-90,width=0.5\textwidth]{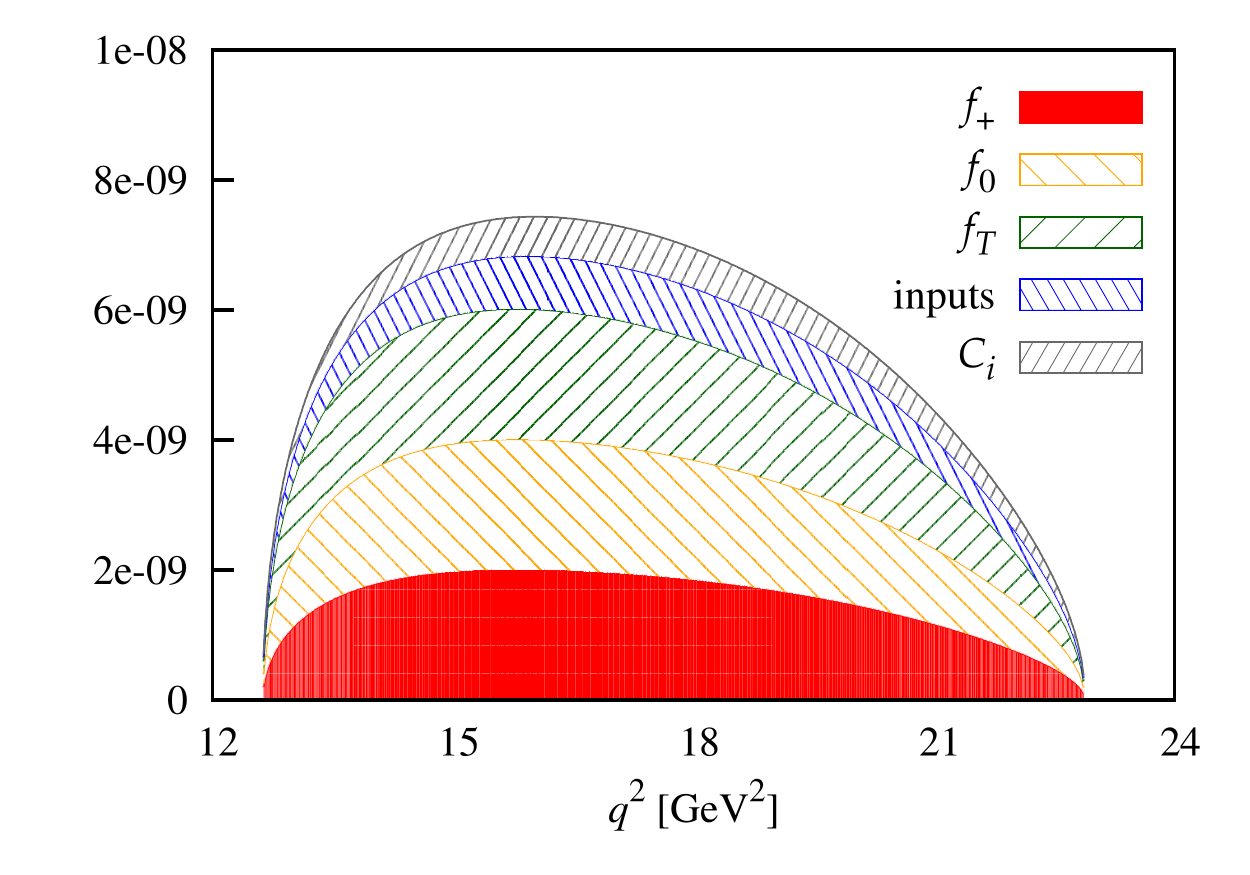}}}
\vspace{0.0in}
\caption{({\it left}) Standard Model differential branching fractions and experiment.  ({\it right}) Form factor, input parameter, and Wilson coefficient ($C_i$) contributions to the error.  The total error is the sum in quadrature of the components.}
\label{fig-dBdq2}
\end{figure*}
Using form factors, determined for the first time from unquenched lattice QCD, we calculate several Standard Model observables that either allow comparison with experiment or make predictions.  
Our form factor results are, within errors, equivalent for $B^0 \to K^0 \ell^+\ell^-$ ($\bar{B}^0 \to \bar{K}^0 \ell^+\ell^-$) and $B^\pm \to K^\pm \ell^+\ell^-$.  The observables we calculate from the form factors introduce additional dependence on $M_B$, $M_K$, and $\tau_B$.  In what follows we calculate isospin averaged values for each observable.
Values for most input parameters are taken from the PDG~\cite{PDG:2012}.  We use $1/\alpha_{\rm EW}=128.957(20)$~\cite{Jegerlehner:2008},  $|V_{tb}V_{ts}^*|=0.0405(8)$~\cite{CKM:2013}, and Wilson coefficients from~\cite{Altmannshofer:2008} with 2\% errors~\cite{ASe}.
Input parameter errors are propagated to errors reported for observables~\cite{gdev}.

Following Ref.~\cite{Becirevic:2012} and restricting ourselves to the Standard Model, the differential decay rate is
\begin{equation}
d\Gamma_\ell / dq^2 = 2a_\ell + 2c_\ell / 3 ,
\end{equation}
where $a_\ell$ and $c_\ell$, defined in~\cite{Bouchard:PRD}, are functions of form factors, Wilson coefficients, and other input parameters.

We convert decay rates into branching fractions using the $B$ meson's mean lifetime, $\mathcal{B}_\ell = \Gamma_\ell \tau_B$.
The resulting differential branching fractions are shown for decay into a generic light dilepton final state in Fig.~\ref{fig-dBFdq2l} and a ditau final state in Fig.~\ref{fig-dBFdq2tau}.  Differential branching fractions for dielectron and dimuon final states are nearly identical and when a generic light dilepton final state is referenced, values are obtained using the average differential branching fraction.  
Figs.~\ref{fig-dBFdq2l_err} and~\ref{fig-dBFdq2tau_err} show error contributions from form factors, input parameters, and Wilson coefficients, denoted $C_i$.  
Uncertainty in the form factors dominates. 
Form factor errors are better controlled in the region of simulated $q^2$.  As a result, differential branching fractions for $B\to K\tau^+\tau^-$ and for light dilepton final states at large $q^2$ are more precisely determined. 

\setlength{\tabcolsep}{0.03in}
\begin{table*}[h!]
\begin{tabular}{ccccccc}
\hline\hline
	\T measurement/  		 		& \multicolumn{6}{c}{ $(q^2_{\rm low}, q^2_{\rm high})\ {\rm GeV}^2$ }  \\
	\T calculation					& $(1, 6)$						& $(4.3,8.68)$  					& $(10.09,12.86)$ 				& $(14.18,16)$					& $(16,18)$	& $(16,q^2_{\rm max})$	  			\\ [0.5ex]
	\hline
	& & & & \\ [-3ex]
	\T BABAR~\cite{Lees:2012}	 	& $1.36^{+0.27}_{-0.24} \pm 0.03 $ 	& $0.94^{+0.20}_{-0.19}\pm0.02$ 	& $0.90^{+0.20}_{-0.19}\pm0.04$  	&$ 0.49^{+0.15}_{-0.14} \pm 0.02 $ 	& -- & $ 0.67^{+0.23}_{-0.21} \pm 0.05 $ 		\\ [0.7ex]
	\T Belle~\cite{Wei:2009}			& $1.36^{+0.23}_{-0.21} \pm 0.08 $ 	& $1.00^{+0.19}_{-0.18}\pm0.06$ 	& $0.55^{+0.16}_{-0.14}\pm0.03$	&$ 0.38^{+0.19}_{-0.12} \pm 0.02 $ 	& -- & $ 0.98^{+0.20}_{-0.18} \pm 0.06 $		\\ [0.7ex]
	\T CDF~\cite{Aaltonen:2011}	 	& $1.29 \pm 0.18 \pm 0.08 $ 		& $1.05 \pm 0.17 \pm 0.07$ 	     	& $0.48\pm0.10\pm0.03$ 			& $ 0.52 \pm 0.09 \pm 0.03 $ 		& -- & $ 0.38 \pm 0.09 \pm 0.02 $ 			\\ [0.7ex]
	\T LHCb~\cite{Aaij:2012}		 	& $ 0.65^{+0.45}_{-0.35} $ 		& $1.22 \pm 0.31$ 				& $0.50^{+0.22}_{-0.19}$ 			& $ 0.20^{+0.13}_{-0.09} $ 		& -- & $ 0.35^{+0.21}_{-0.14} $ 			\\ [0.7ex]
	\T LHCb~\cite{Aaij:2012b}	 	& $ 1.21\pm0.09\pm0.07 $ 		& $1.00\pm0.07\pm0.04$ 			& $0.57\pm0.05\pm0.02$ 			& $0.38\pm0.04\pm0.02$ 			& $0.35\pm0.04\pm0.02$ & --				\\ [0.7ex]
	\hline
	\T this work				 	& $1.81 \pm 0.61 $ 				& $ 1.65 \pm 0.42 $				& $0.87\pm 0.13$ 				& $ 0.442\pm 0.051 $			& $0.391\pm 0.042$ & $ 0.797\pm 0.082 $ 				\\ [0.7ex]
	\T Ref.~\cite{Altmannsofer:2012}	& $ 1.29 \pm 0.30 $				& --							& --							& $ 0.43 \pm 0.10 $				& -- & $ 0.86 \pm 0.20 $					\\ [0.7ex]
	\T Ref.~\cite{Bobeth:2013} 		& $ 1.63^{+0.56}_{-0.27} $		& $ 1.38^{+0.51}_{-0.25} $		& --							& $ 0.340^{+0.179}_{-0.083} $		& $0.309^{+0.176}_{-0.081}$ & $ 0.634^{+0.382}_{-0.175}$			\\ [0.7ex]
	\T Ref.~\cite{Khodjamirian:2013} 	& $1.76^{+0.60}_{-0.23}$			& $1.39^{+0.53}_{-0.22}$			& --							& --							& -- & -- 								\\ [0.7ex]
\hline\hline
\end{tabular}\caption{Comparison of experiment and theory for $ 10^7 \mathcal{B}_\ell(q^2_{\rm low}, q^2_{\rm high})$, with $\ell = e, \mu$, for various ranges of integration.  BABAR~\cite{Lees:2012} uses slightly different $q^2$ bins:   $(4.3,8.12)$, $(10.11,12.89)$, and $(14.21,16)$.
CDF~\cite{Aaltonen:2011} measurements are isospin averaged and for dimuon final states.
LHCb measurements are for $B^0\to K^0 \mu^+ \mu^-$~\cite{Aaij:2012} and $B^+\to K^+ \mu^+\mu^-$~\cite{Aaij:2012b}.  
Quoted values from Bobeth {\it et al.}~\cite{Bobeth:2013} are for $\bar{B}^0 \to \bar{K}^0 \ell^+\ell^-$.}
\label{tab-BFvexpt}
\end{table*}

\setlength{\tabcolsep}{0.05in}
\begin{table*}[t!]
\vspace{0.27in}
\begin{tabular}{ccccccc}
\hline\hline
	\T   							& \multicolumn{6}{c}{ $(q^2_{\rm low}, q^2_{\rm high})\ {\rm GeV}^2$ }  																				\\
	\T observable					& $(1, 6)$						& $(4.3,8.68)$					& $(10.09,12.86)$		& $(14.18,16)$ 				& $(16,18)$ 			& $(16,q^2_{\rm max})$	  	\\ [0.5ex]
	\hline
	\T							& 							& 							& 					& 						& $$ 					& 						\\ [-3ex]
	\T $10^3(R^\mu_e-1)$ 			& $0.74\pm0.35$ 				& $0.89\pm0.25$				& $1.35\pm0.23$		& $1.98\pm0.22$ 			& $2.56\pm0.23$ 		& $3.86\pm0.29$			\\ [0.1ex]
	\T Ref.~\cite{Bobeth:2007}		& $0.31^{+0.10}_{-0.07}$			& --							& --					& --						& -- 					& --						\\ [2.0ex]
	\T $10^6  F_H^e$				& $0.577\pm0.010$ 				& $0.2722\pm0.0054$			& $0.1694\pm0.0053$	& $0.1506\pm0.0052$ 		& $0.1525\pm0.0055$ 	& $0.1766\pm0.0068$ 		\\ [2.0ex]
	\T $10^2 F_H^\mu$				& $2.441\pm0.043$ 				& $1.158\pm0.023$				& $0.722\pm0.022$		& $0.642\pm0.022$			& $0.649\pm0.023$ 		& $0.751\pm0.029$ 			\\ [0.1ex]
	\T Ref.~\cite{Bobeth:2013}		& $2.54^{+0.20}_{-0.36}$			& $1.24^{+0.12}_{-0.20}$			& --					& $0.704^{+0.147}_{-0.196}$	& $0.318^{+0.201}_{-0.092}$ & $0.775^{+0.210}_{-0.254}$ 	\\ [0.1ex]
	\T LHCb~\cite{Aaij:2012b}		& $5.0^{+8.0\ +4.0}_{-5.0\ -2.0}$	& $4.0^{+10.0\ +6.0}_{-4.0\ \ - 4.0}$	& $11.0^{+20.0\ +2.0}_{-8.0\ \ +1.0}$	 & $8.0^{+28.0\ +2.0}_{-8.0\ \ +1.0}$	& $18.0^{+22.0\ +1.0}_{-14.0\ -4.0}$ & -- 		\\ [0.7ex]
\hline\hline
\end{tabular}\caption{Binned light dilepton observables compared with LHCb~\cite{Aaij:2012b} and other selected results~\cite{Bobeth:2013, Bobeth:2007}.}
\label{tab-RFbins}
\end{table*}
\setlength{\tabcolsep}{0.1in}
\begin{table*}[t]
\begin{tabular}{ccccc}
\hline\hline
	\T   												& \multicolumn{4}{c}{ $(q^2_{\rm low}, q^2_{\rm high})\ {\rm GeV}^2$ }  \\
	\T observable										& $(14.18,q^2_{\rm max})$	& $(14.18,16)$ 			& $(16,18)$			& $(16,q^2_{\rm max})$	  \\ [0.5ex]
	\hline
													& 						& 					& 					& 					\\ [-3ex]
	\T $R^\tau_\mu(q^2_{\rm low},q^2_{\rm high})$			& $ 1.158 \pm 0.039 $		& $0.790\pm0.025$ 		& $1.055\pm0.033$ 		& $1.361\pm0.046$		\\ [0.1ex]
	\T $R^\tau_e(q^2_{\rm low},q^2_{\rm high})$				& $ 1.161 \pm 0.040 $ 		& $0.792\pm0.025$ 		& $1.058\pm0.034$ 		& $1.367\pm0.047$ 		\\ [0.1ex]
	\T $R^\tau_\ell(q^2_{\rm low},q^2_{\rm high})$				& $ 1.159 \pm 0.040 $ 		& $0.791\pm0.025$ 		& $1.056\pm0.033$ 		& $1.364\pm0.046$ 		\\ [2.0ex]
	\T $F_H^\tau(q^2_{\rm low},q^2_{\rm high})$				& $0.8856\pm0.0037$		& $0.9176\pm0.0026$	& $0.8784\pm0.0038$ 	& $0.8753\pm0.0042$	\\ [0.1ex]
	\T Ref.~\cite{Bobeth:2011}							& $0.890^{+0.033}_{-0.045}$	& -- 					& --					& --					\\ [2.0ex]
	\T $10^7\mathcal{B}_\tau(q^2_{\rm low},q^2_{\rm high})$		& $1.44\pm0.15$			& $0.349\pm0.040$		& $0.413\pm0.044$ 		& $1.09\pm0.11$		\\ [0.1ex]
	\T Ref.~\cite{Bobeth:2011}							& $1.26^{+0.41}_{-0.23}$		& --					& --					& --			\\ [0.7ex]
\hline\hline
\end{tabular}\caption{Binned ditau final state observables.  We compare with results for the flat term and branching fraction from Bobeth {\it et al.}~\cite{Bobeth:2011}.}
\label{tab-RFtaubins}
\end{table*}

Integrating the differential branching fractions over $q^2$ bins defined by $(q^2_{\rm low}, q^2_{\rm high})$ permits direct comparison with experiment,
\begin{equation}
\mathcal{B}_\ell(q^2_{\rm low}, q^2_{\rm high}) \equiv \int_{q^2_{\rm low}}^{q^2_{\rm high}} dq^2\ d\mathcal{B}_\ell / dq^2\ .
\end{equation}
Integrating over the full kinematic range yields the total branching fractions
\begin{eqnarray}
10^7  \mathcal{B}_e(4m_e^2, q^2_{\rm max}) &=& 6.14\pm1.33, \nonumber \\
10^7  \mathcal{B}_\mu(4m_\mu^2, q^2_{\rm max}) &=& 6.12\pm1.32, \nonumber \\
10^7 \mathcal{B}_\tau(14.18\ {\rm GeV}^2, q^2_{\rm max}) &=& 1.44\pm0.15,
\end{eqnarray}
where $q^2_{\rm max}=(M_B-M_K)^2$.  For the ditau final state we begin the integration at $14.18\ {\rm GeV}^2$ to account for the experimentally vetoed $\psi(2S)$ region.
A detailed comparison of our Standard Model branching fraction results with experiment, and other calculations, is given in Table~\ref{tab-BFvexpt}.  The results of Altmannshofer and Straub~\cite{Altmannsofer:2012} use form factors from Ref.~\cite{Bharucha:2010}, in which quenched lattice~\cite{Al-Haydari:2009} and light cone sum rule~\cite{Ball:2005} results are combined. 
The results of Bobeth {\it et al.}~\cite{Bobeth:2013} use form factors obtained from light cone sum rules in Ref.~\cite{Khodjamirian:2010} and extrapolated to large $q^2$ via $z$ expansion.

The ratio of dimuon and dielectron branching fractions
\begin{equation}
R^\mu_e(q^2_{\rm low}, q^2_{\rm high}) \equiv \frac{ \int_{q^2_{\rm low}}^{q^2_{\rm high}} dq^2\ d\mathcal{B}_\mu / dq^2 }{ \int_{q^2_{\rm low}}^{q^2_{\rm high}} dq^2\ d\mathcal{B}_e / dq^2},
\end{equation}
is a potentially sensitive probe of new physics~\cite{Hiller:2004}, though measurements thus far~\cite{Wei:2009, Lees:2012} have been consistent with the Standard Model.
We extend the ratio to ditau final states, where new physics contributions may be even larger~\cite{Yan:2000} and find
\begin{eqnarray}
R^\mu_e(4m_\mu^2, q^2_{\rm max}) &=& 1.00023(63), \\
R^\tau_\mu(14.18\ {\rm GeV}^2, q^2_{\rm max}) &=& 1.158(39), \\
R^\tau_e(14.18\ {\rm GeV}^2, q^2_{\rm max}) &=& 1.161(40), \\
R^\tau_\ell(14.18\ {\rm GeV}^2, q^2_{\rm max}) &=& 1.159(40).
\end{eqnarray}
Correlations among form factors are accounted for in the calculation of the ratios.
We give values of the branching fraction ratios in different $q^2$ bins in Tables~\ref{tab-RFbins} and~\ref{tab-RFtaubins}.

The angular distribution of the differential decay rate is given by
\begin{equation}
\frac{1}{\Gamma_\ell}\frac{d \Gamma_\ell}{d \cos\theta_\ell} = \frac{1}{2}F_H^\ell + A_{FB}^\ell \cos\theta_\ell + \frac{3}{4}(1-F_H^\ell)(1-\cos^2\theta_\ell),
\end{equation}
where $\theta_\ell$ is the angle between the $B$ and $\ell^-$ as measured in the dilepton rest frame.  The ``flat term" $F_H^\ell$, introduced by Bobeth {\it et al.}~\cite{Bobeth:2007}, is suppressed by $m_\ell^2$ in the Standard Model and is potentially sensitive to new physics~\cite{Bobeth:2013, Becirevic:2012}.  The ``forward-backward asymmetry" $A_{FB}^\ell$ is zero in the Standard Model (up to negligible QED contributions~\cite{Demir:2000, Bobeth:2007}) so is also a sensitive probe of new physics.  The flat term~\cite{Bobeth:2007}
\begin{equation}
F_H^\ell(q^2_{\rm low}, q^2_{\rm high}) = \frac{ \int_{q^2_{\rm low}}^{q^2_{\rm high}} dq^2\ (a_\ell + c_\ell) }{ \int_{q^2_{\rm low}}^{q^2_{\rm high}} dq^2\ (a_\ell + c_\ell / 3) }
\end{equation}
is constructed as a ratio to reduce uncertainties.  Evaluated in experimentally motivated $q^2$ bins, values for $F_H^{e,\mu, \tau}$ are given in Tables~\ref{tab-RFbins} and~\ref{tab-RFtaubins}.

\section{ Summary and Outlook }
Employing unquenched lattice QCD form factors for the rare decay $B\to K\ell^+\ell^-$~\cite{Bouchard:PRD}, we calculate the first model-independent Standard Model predictions for:  
differential branching fractions; 
branching fractions integrated over experimentally motivated $q^2$ bins; 
ratios of branching fractions potentially sensitive to new physics; 
and the flat term in the angular distribution of the differential decay rate.  Where available, we compare with experiment and previous calculations.  For $q^2 \gtrsim 10\ {\rm GeV}^2$ our results are more precise than previous Standard Model predictions.  
For all $q^2$ our results are consistent with previous calculations and experiment.

Predictions for observables involving the ditau final state are particularly precise and potentially sensitive to new physics.  Given this combination, measurements of $\mathcal{B}_\tau$, $R^\tau_\ell$, or $F_H^\tau$ by experimentalists would be particularly interesting and welcome.

\section{ Acknowledgements}
This research was supported by the DOE and NSF.
We thank the MILC collaboration for making their asqtad $N_f=2+1$ gauge field configurations available.  
Computations were carried out at the Ohio Supercomputer Center and on facilities of the USQCD collaboration funded by the Office of Science of the U.S. DOE.


\begin{thebibliography}{99}

\bibitem{Becirevic:2012}
   D.~Be\v{c}irevi\'{c}, N.~Ko\v{s}nik, F.~Mescia, and E.~Schneider,
   Phys. Rev. D {\bf 86}, 034034 (2012)
   [\href{http://arxiv.org/abs/1205.5811}{1205.5811 [hep-ph]}].

\bibitem{Bobeth:2011}
   C.~Bobeth, G.~Hiller, D.~van~Dyk, and C.~Wacker,
   {\it JHEP} {\bf 01} (2012) 107
   [\href{http://arxiv.org/abs/1111.2558}{1111.2558 [hep-ph]}].
   
\bibitem{Beaujean:2012}
   F.~Beaujean, C.~Bobeth, D.~van~Dyk, and C.~Wacker,
   {\it JHEP} {\bf 08} (2012) 030
   [\href{http://arxiv.org/abs/1205.1838}{1205.1838 [hep-ph]}].

\bibitem{Altmannsofer:2012}
   W.~Altmannshofer and D.~M.~Straub,
   {\it JHEP} {\bf 08} (2012) 121
   [\href{http://arxiv.org/abs/1206.0273}{1206.0273 [hep-ph]}].

\bibitem{Bobeth:2013}
   C.~Bobeth, G.~Hiller, and D.~van~Dyk,
   Phys. Rev. D {\bf 87}, 034016 (2013)
   [\href{http://arxiv.org/abs/1212.2321}{1212.2321 [hep-ph]}].
   
\bibitem{Ball:2005}
    P.~Ball and R.~Zwicky,
    Phys. Rev. D {\bf 71}, 014029 (2005)
    [\href{http://xxx.lanl.gov/abs/hep-ph/0412079}{hep-ph/0412079}].

\bibitem{Khodjamirian:2010}
   A.~Khodjamirian, Th.~Mannel, A.~A.~Pivovarov, and Y.-M.~Wang,
   {\it JHEP} {\bf 09} (2010) 089
   [\href{http://arxiv.org/abs/arXiv:1006.4945}{1006.4945 [hep-ph]}].
   
\bibitem{Khodjamirian:2013}
   A.~Khodjamirian, Th.~Mannel, and Y.-M.~Wang,
   {\it JHEP} {\bf 02} (2013) 010
   [\href{http://arxiv.org/pdf/1211.0234v2.pdf}{1211.0234 [hep-ph]}].

\bibitem{quench}
  Including the effects of virtual quark-antiquark pairs in lattice QCD simulations, through the determinant of the Dirac operator, is computationally demanding.  Such effects were omitted in early lattice calculations by setting the determinant to one.  The error associated with this approximation, called quenching, is uncontrolled and difficult to estimate.  Only in the past decade has it become possible to include these effects and perform realistic lattice calculations with controlled systematic errors.

\bibitem{Becirevic:2000}
   D.~Be\v{c}irevi\'{c} and A.~B.~Kaidalov,
   Phys. Lett. B {\bf 478}, 417 (2000)
   [\href{http://arxiv.org/abs/hep-ph/9904490}{hep-ph/9904490}].
   
 \bibitem{Wei:2009}
   J.-~T.~Wei {\it et al.}
   (Belle),
   Phys. Rev. Lett. {\bf 103}, 171801 (2009)
   [\href{http://arxiv.org/abs/arXiv:0904.0770}{0904.0770 [hep-ex]}].
   
\bibitem{Lees:2012}
   J.~P.~Lees {\it et al.}
   (BABAR),
   Phys. Rev. D {\bf 86}, 032012 (2012)
   [\href{http://arxiv.org/abs/arXiv:1204.3933}{1204.3933 [hep-ex]}].

\bibitem{Aaltonen:2011}
   T.~Aaltonen {\it et al.}
   (CDF),
   Phys. Rev. Lett. {\bf 107}, 201802 (2011)
   [\href{http://arxiv.org/abs/arXiv:1107.3753}{1107.3753 [hep-ex]}].

\bibitem{Aaij:2012}
   R.~Aaij {\it et al.}
   (LHCb),
   {\it JHEP} {\bf 07} (2012) 133
   [\href{http://arxiv.org/abs/1205.3422}{1205.3422 [hep-ex]}].
   
\bibitem{Aaij:2012b}
   R.~Aaij {\it et al.}
   (LHCb),
   {\it JHEP} {\bf 02} (2013) 105
   [\href{http://arxiv.org/abs/1209.4284}{1209.4284 [hep-ex]}].
   
\bibitem{Liu:2011}
   Z.~Liu, S.~Meinel, A.~Hart, R.~Horgan, E.~M\"{u}ller, and M.~Wingate,
   \href{http://arxiv.org/pdf/1101.2726v1.pdf}{1101.2726 [hep-ph]}.

\bibitem{Zhou:2012}
   R.~Zhou, S.~Gottlieb, J.~A.~Bailey, D.~Du, A.~X.~El-Khadra, R.~D.~Jain, A.~S.~Kronfeld, R.~S.~Van~de~Water, Y.~Liu, and Y.~Meurice
   (Fermilab Lattice and MILC),
   [\href{http://arxiv.org/abs/1211.1390}{1211.1390 [hep-lat]}].
   
\bibitem{Bouchard:PRD}
   C.~M.~Bouchard, G.~P.~Lepage, C.~Monahan, H.~Na, and J.~Shigemitsu
   (HPQCD),
   Phys. Rev. D {\bf 88}, 054509 (2013);
   Phys. Rev. D {\bf 88}, 079901(E) (2013)
   [\href{http://arxiv.org/pdf/1306.2384v1.pdf}{1306.2384 [hep-lat]}].
   
\bibitem{Bazavov:2010}
   A.~Bazavov, C.~Bernard, C.~DeTar, S.~Gottlieb, U.~M.~Heller, J.~E.~Hetrick, J.~Laiho, L.~Levkova, P.~B.~Mackenzie, M.~B.~Oktay, R.~Sugar, D.~Toussaint, and R.~S.~Van~de~Water
   (MILC),
   Rev. Mod. Phys. {\bf 82}, 1349 (2010)
   [\href{http://arxiv.org/abs/0903.3598}{0903.3598 [hep-lat]}].

\bibitem{Lepage:1992}
   G.~P.~Lepage, L.~Magnea, C.~Nakhleh, U.~Magnea, and K.~Hornbostel
   (HPQCD),
   Phys. Rev. D {\bf 46}, 4052 (1992)
   [\href{http://arxiv.org/abs/hep-lat/9205007}{hep-lat/9205007}].

\bibitem{Na:2012}
   H.~Na, C.~J.~Monahan, C.~T.~H.~Davies, R.~Horgan, G.~P.~Lepage and J.~Shigemitsu
   (HPQCD),
   Phys. Rev. D {\bf 86}, 034506 (2012)
   [\href{http://arxiv.org/abs/1202.4914}{1202.4914 [hep-lat]}].
   
\bibitem{Na:2010}
   H.~Na, C.~T.~H.~Davies, E.~Follana, G.~P.~Lepage, and J.~Shigemitsu
   (HPQCD),
   Phys. Rev. D {\bf 82}, 114506 (2010)
   [\href{http://arxiv.org/abs/1008.4562}{1008.4562 [hep-lat]}].

\bibitem{Na:2011}
   H.~Na, C.~T.~H.~Davies, E.~Follana, J.~Koponen, G.~P.~Lepage, and J.~Shigemitsu
   (HPQCD),
   Phys. Rev. D {\bf 84}, 114505 (2011)
   [\href{http://arxiv.org/abs/1109.1501}{1109.1501 [hep-lat]}].

   


\bibitem{Follana:2007}
   E.~Follana, Q.~Mason, C.~Davies, K.~Hornbostel, G.~P.~Lepage, J.~Shigemitsu, H.~Trottier, and K.~Wong
   (HPQCD),
   Phys. Rev. D {\bf 75}, 054502 (2007)
   [\href{http://arxiv.org/abs/hep-lat/0610092}{hep-lat/0610092}].
   
\bibitem{Lepage:2002}
   G.~P.~Lepage, B.~Clark, C.~T.~H.~Davies, K.~Hornbostel, P.~B.~Mackenzie, C.~Morningstar and H.~Trottier,
   Nucl. Phys. Proc. Suppl.  {\bf 106}, 12 (2002)
   [\href{http://arxiv.org/abs/hep-lat/0110175}{hep-lat/0110175}].

\bibitem{Monahan:2013}
   C.~Monahan, J.~Shigemitsu, and R.~Horgan
   (HPQCD),
   Phys. Rev. D {\bf 87}, 034017 (2013)
   [\href{http://arxiv.org/abs/arXiv:1211.6966}{1211.6966 [hep-lat]}].



   
\bibitem{Aubin:2007}
   C.~Aubin and C.~Bernard,
   Phys. Rev. D {\bf 76}, 014002 (2007)
   [\href{http://arxiv.org/abs/0704.0795}{0704.0795 [hep-lat]}].
   
   
   
\bibitem{Boyd:1996}  
   C.~G.~Boyd, B.~Grinstein, and R.~F.~Lebed,
   Nucl. Phys. B {\bf 461}, 493 (1996)
   [\href{http://arxiv.org/abs/hep-ph/9508211}{hep-ph/9508211}].

\bibitem{Arnesen:2005}
   M.~C.~Arnesen, B.~Grinstein, I.~Z.~Rothstein, and I.~W.~Stewart,
   Phys. Rev. Lett. {\bf 95}, 071802 (2005)
   [\href{http://arxiv.org/abs/hep-ph/0504209}{hep-ph/0504209}].
   
\bibitem{Bourrely:2010}
   C.~Bourrely, L.~Lellouch, and I.~Caprini,
   Phys. Rev. D {\bf 79}, 013008 (2009); 
   Phys. Rev. D {\bf 82}, 099902(E) (2010)
   [\href{http://arxiv.org/abs/0807.2722}{0807.2722 [hep-ph]}].
   
\bibitem{Davies:2008}
   C.~T.~H.~Davies,  K.~Hornbostel, I.~D.~Kendall, G.~P.~Lepage, C.~McNeile, J.~Shigemitsu, and H.~Trottier
   (HPQCD),
   Phys. Rev. D {\bf 78} 114507 (2008)
   [\href{http://arxiv.org/abs/0807.1687}{0807.1687 [hep-lat]}].
   
\bibitem{PDG:2012}
    J.~Beringer {\it et al.}
    (Particle Data Group),
    Phys. Rev. D {\bf 86}, 010001 (2012) and 2013 partial update for the 2014 edition
    [\href{http://pdg.lbl.gov/}{pdg.lbl.gov}].
    
\bibitem{Jegerlehner:2008}
   F.~Jegerlehner,
   Nucl. Phys. Proc. Suppl. {\bf 181-182}, 135-140 (2008)
   [\href{http://arxiv.org/abs/arXiv:0807.4206}{0807.4206 [hep-ph]}].
    
 \bibitem{CKM:2013}
    J.~Charles {\it et al.}
    (CKMfitter Group), 
    Eur. Phys. J. {\bf C41}, 1-131 (2005) 
    [\href{http://arxiv.org/abs/hep-ph/0406184}{hep-ph/0406184}], 
    updated results and plots available at: \href{http://ckmfitter.in2p3.fr}{ckmfitter.in2p3.fr}.   

\bibitem{Altmannshofer:2008}
   W.~Altmannshofer, P.~Ball, A.~Bharucha, A.~J.~Buras, D.~M.~Straub, and M.~Wick,
    {\it JHEP} {\bf 01} (2009) 019
   [\href{http://arxiv.org/abs/0811.1214}{0811.1214 [hep-ph]}].
   
\bibitem{ASe}
   W.~Altmannshofer and D.~M.~Straub (private communication).
   
\bibitem{gdev}
   This is accomplished with the use of {\tt gvar} data types in the {\tt lsqfit} Python package, available at \href{http://www.physics.gla.ac.uk/HPQCD/}{www.physics.gla.ac.uk/HPQCD}.

\bibitem{Bharucha:2010}
   A.~Bharucha, T.~Feldmann, and M.~Wick
   {\it JHEP} {\bf 09} (2010) 090
   [\href{http://arxiv.org/abs/1004.3249}{1004.3249 [hep-ph]}].
   
\bibitem{Al-Haydari:2009}
   A.~Al-Haydari, A.~Ali~Khan, V.~M.~Braun, S.~Collins, M.~G\"{o}ckeler, G.~N.~Lacagnina, M.~Panero, A.~Sch\"{a}fer, G.~Schierholz
   (QCDSF),
   Eur. Phys. J. A {\bf 43}, 107-120 (2010)
   [\href{http://xxx.lanl.gov/abs/0903.1664}{0903.1664 [hep-lat]}].   
   
\bibitem{Hiller:2004}
   G.~Hiller and F.~Kr\"{u}ger,
   Phys. Rev. D {\bf 69}, 074020 (2004)
   [\href{http://arxiv.org/abs/hep-ph/0310219}{hep-ph/0310219}].

\bibitem{Yan:2000}
   Q.-S.~Yan, C.-S.~Huang, W.~Liao, and S.-H.~Zhu,
   Phys. Rev. D {\bf 62}, 094023 (2000)
   [\href{http://arxiv.org/abs/hep-ph/0004262}{hep-ph/0004262}].
   
\bibitem{Bobeth:2007}
   C.~Bobeth, G.~Hiller, and G.~Piranishvili,
   {\it JHEP} {\bf 12} (2007) 040
   [\href{http://arxiv.org/abs/0709.4174}{0709.4174 [hep-ph]}].
    
\bibitem{Demir:2000}
   D.~A.~Demir, K.~A.~Olive, and M.~B.~Voloshin,
   Phys. Rev. D {\bf 66}, 034015 (2002)
   [\href{http://arxiv.org/abs/hep-ph/0204119}{hep-ph/0204119}].
   
   
   
%
%
%
%
%
%
%
%
%
%
%
%
%
%
%
%


\end{thebibliography}


\end{document}